\documentclass[aps,prd,preprint,superscriptaddress,showpacs,nofootinbib]{revtex4-1}

\usepackage{graphicx}
\usepackage{amssymb}

\begin{document}

\title{Energy versus centrality dependence of the jet quenching parameter $\hat q$ at RHIC and LHC: a new puzzle?}

\author{Carlota Andr\'es}
\email{carlota.andres@usc.es}
\affiliation{
        Instituto Galego de F\'\i sica de Altas Enerx\'\i as IGFAE, Universidade de Santiago de Compostela, E-15782 Santiago de Compostela (Galicia-Spain)}

\author{N\'estor Armesto}
\email{nestor.armesto@usc.es}
\affiliation{
        Instituto Galego de F\'\i sica de Altas Enerx\'\i as IGFAE, Universidade de Santiago de Compostela, E-15782 Santiago de Compostela (Galicia-Spain)}

\author{Matthew Luzum}
\email{mluzum@usp.br}
\affiliation{
        Instituto Galego de F\'\i sica de Altas Enerx\'\i as IGFAE, Universidade de Santiago de Compostela, E-15782 Santiago de Compostela (Galicia-Spain)}
\affiliation{Instituto de F\'\i sica - Universidade de S\~ao Paulo, Rua do Mat\~ao Travessa R,
no. 187, 05508-090, Cidade Universit\'aria, S\~ao Paulo (Brasil)}

\author{Carlos A. Salgado}
\email{carlos.salgado@usc.es}
\affiliation{
        Instituto Galego de F\'\i sica de Altas Enerx\'\i as IGFAE, Universidade de Santiago de Compostela, E-15782 Santiago de Compostela (Galicia-Spain)}

\author{P\'\i a Zurita}
\email{pia.zurita@usc.es}
\affiliation{
        Instituto Galego de F\'\i sica de Altas Enerx\'\i as IGFAE, Universidade de Santiago de Compostela, E-15782 Santiago de Compostela (Galicia-Spain)}

\date{\today}

\begin{abstract}
The central goal of jet quenching studies in high-energy nuclear collisions  is the characterization of those QCD medium properties that are accessible by these probes. Most of the discussion in the last years has been focused on  the determination of the jet quenching parameter, $\hat q$. We present here an extraction of this parameter using data of inclusive particle suppression at RHIC and LHC energies for different centralities. Our approach consists of fitting a $K$ factor that quantifies the departure of this parameter from an ideal estimate, $K\equiv \hat q/(2\epsilon^{3/4})$, where $\hat q$ is determined by the local medium quantities as provided by hydrodynamical calculations. We find that this $K$ factor is larger at RHIC than at the LHC, as obtained already in previous analyses, but, surprisingly, it is almost independent of the centrality of the collision. Taken at face value, the $K$ factor would not depend on the local properties of the medium as energy density or temperature, but on global collision quantities such as the center of mass energy. This is a very intriguing, unexpected possibility for which we cannot yet provide a clear interpretation. We also comment on the limitations of the formalism that may affect this conclusion.
\end{abstract}



\maketitle

\section{Introduction}

Jet quenching --- the suppression of high-energy particles and jets in nucleus-nucleus collisions relative to the expectation from a superposition of nucleon-nucleon ones ---  is one of the best available tools to characterize the properties of the medium created in collisions of heavy nuclei at high energies.  See the recent reviews \cite{Majumder:2010qh,Mehtar-Tani:2013pia,Blaizot:2015lma,Qin:2015srf}. A huge number of experimental data is available and has been phenomenologically studied in the last 15 years, starting with the suppression of inclusive particle production at high transverse momentum at the Relativistic Heavy-Ion Collider (RHIC) and now at the Large Hadron Collider (LHC). At present, a large experimental effort is devoted to the study of reconstructed jets, where more differential analyses are possible for a more precise characterization of medium properties. For reference to the experimental works,  see e.g.~the reviews \cite{Norbeck:2014loa,Armesto:2015ioy}.

On the theory side, much progress has also been achieved recently on exploiting and improving the standard picture of radiative and collisional energy loss from Refs. \cite{Baier:1996kr,Baier:1996sk,Zakharov:1997uu,Gyulassy:2000er,Wiedemann:2000za,Wang:2001ifa,Arnold:2002ja}. The different implementations have been critically revised in \cite{Armesto:2011ht} and employed to extract medium parameters in \cite{Burke:2013yra,Liu:2015vna}, to quote some recent references. Monte Carlo implementations have been developed \cite{Lokhtin:2008xi,Renk:2008pp,Armesto:2009fj,Armesto:2009ab,Schenke:2009gb,Zapp:2012ak,Majumder:2013re}, and improvements on the kinematics have been computed in \cite{Apolinario:2012vy,Blaizot:2012fh,Fickinger:2013xwa,Apolinario:2014csa}. The effect of the interplay between collisional and radiative energy loss on jet yields and shapes has been analyzed in \cite{CasalderreySolana:2010eh,Qin:2010mn,Qin:2012fua} (see also Ref. \cite{Wicks:2005gt}) and a combination of weak and strong coupling approaches has been studied in \cite{Casalderrey-Solana:2014bpa}. Finally, much theoretical work has been devoted to the development of a complete picture of in-medium parton branching, both in the frame of soft-collinear effective theory \cite{Fickinger:2013xwa,Ovanesyan:2011xy,Kang:2014xsa} and using the QCD antenna as a setup \cite{MehtarTani:2010ma,MehtarTani:2011tz,MehtarTani:2011jw,MehtarTani:2011gf,MehtarTani:2012cy,CasalderreySolana:2012ef,Armesto:2013fca,CasalderreySolana:2011rz,Arleo:2010rb,Blaizot:2013vha}.

The final goal of the jet quenching studies is to extract medium parameters which characterize the QCD matter formed in high-energy nuclear collisions. In this paper we present  an analysis of RHIC and LHC data on the nuclear modification factor $R_{AA}$  for inclusive particle production at high transverse momentum using the formalism of the {\it quenching weights} proposed in \cite{Baier:2001yt,Salgado:2002cd,Salgado:2003gb}. The main result of this paper is an extraction of the value for the jet quenching parameter $\hat q$ in a method which has been well tested, is easy to implement and interplay with different hydrodynamical models for the medium (as done previously by some of the authors of the present paper in Ref. \cite{Armesto:2009zi}, see also \cite{Burke:2013yra,Renk:2011gj,Bass:2008rv}), and provides a good description of the experimental data, as we will show. Despite the limitations of the formalism whose applicability is restricted to describing leading particle production in a jet, we argue that the main conclusions of the analysis are solid and somehow unexpected. 

Our approach here is to define a local transport coefficient, $\hat q$, which is solely determined by the local (in position and time) medium quantities as extracted from hydrodynamical models, in particular the energy density. So, we define the jet quenching parameter as $\hat q =K2\epsilon^{3/4}$, motivated by the ideal estimate $\hat q_{\rm ideal}\sim 2\epsilon^{3/4}$ \cite{Baier:2002tc}, and fit the values of $K$. We do not impose any particular dependence of this $K$-factor on energy, centrality, temperature, etc. On the contrary, $K$ is the only free parameter in the fit of $R_{AA}$ for each centrality at RHIC and LHC energies. Our main findings are that this $K$-factor is $\sim 2 - 3$ times larger for RHIC than for the LHC (larger values at RHIC than at the LHC have been found before \cite{Horowitz:2011gd}) and, unexpectedly, this $K$-factor does not seem to depend on the medium parameters, e.g., the temperature, but instead on the center of mass energy of the collision. Indeed, we find these $K$-factors to be basically independent of centrality both for RHIC and the LHC. Were the $K$-factor determined, say, by temperature, then the most central RHIC collisions should present a value similar to semi-peripheral LHC data. This is not the case. We have performed the study with several, quite different, hydrodynamical profiles and these conclusions do not depend on the profile we use, although the values of $K$ do present some dependences and, interestingly, they dramatically change for some different assumptions for the dynamics at initial times before the starting of hydrodynamical evolution.

We have no clear interpretation of this finding. We comment on the different theoretical limitations of the procedure which could affect this result. A more detailed study of these limitations is not easy with present theoretical tools but we expect that it will be possible in the near future. Taken at face value, if our result is not due to a limitation of the technique, it would indicate that the properties of the QCD media produced at RHIC and LHC are different in what concerns the jet quenching process. Some possibilities could be related with the initial stages of the collision, the presence of quantities related with the total energy of the collision (relative differences between local and global bulk/thermodynamical properties, the presence of a magnetic field, etc.) or others. At this point, however, we do not speculate with these possibilities and present our findings as they are obtained. It would be important to check our conclusion with other jet quenching implementations.

The rest of the paper is organized as follows: we present the formalism of energy loss in Section \ref{sect:formalism}; in Section \ref{sect:hydroimp} we describe how the energy loss  is interfaced with the profiles taken from hydrodynamical models; in Section \ref{sect:hydro} we briefly comment on the details of the three hydrodynamical models used in our analysis; in Section \ref{sect:results} we present the results of the fits for $R_{AA}$ for both RHIC and LHC; finally, we present our conclusions.

\section{Energy loss formalism}
\label{sect:formalism}

The building block of the energy loss model is the one-gluon inclusive energy spectrum of gluons with energy $\omega$ emitted off a highly energetic colored particle traversing a QCD medium, $\omega dN/d\omega$ \cite{Salgado:2003gb,Armesto:2003jh}. In order to compute the energy loss, $\Delta E$, the distribution $P(\Delta E)$ including all possible contributions, not only from single gluon emission but also from two, three, etc., is needed. The way in which the spectrum, $dN/d\omega$, is related to the distribution of lost energy $P(\Delta E)$ is not exactly known: as usual in perturbation theory, an all-order computation is not possible. Starting with the proposal of independent gluon emission approximation  \cite{Baier:2001yt}, several groups have just iterated the one-gluon inclusive in an independent manner \cite{Gyulassy:2001nm,Salgado:2003gb,Armesto:2005iq}, which has been the most standard procedure to deal with the problem in the last years. An implementation of these ideas, the {\it quenching weights}, was worked out in Ref. \cite{Salgado:2003gb} and will be used in this paper. Note that these quenching weights can be obtained from an iterative solution of DGLAP evolution for medium-modified fragmentations functions in the soft limit \cite{Armesto:2007dt}.

An attempt to include the single-inclusive emission in the form of a kernel in the more sophisticated manner of a rate equation was proposed in \cite{Jeon:2003gi,Turbide:2005fk} and incorporated  into what is known as the AMY framework. This rate equation still assumed with no proof that subsequent emissions are independent. It is with a more developed study of the role of coherence in jet quenching \cite{MehtarTani:2010ma,CasalderreySolana:2011rz} that one can prove that in the limit of gluon formation times much smaller than the medium size, $\tau_{\rm form}\ll L$, a resummation is possible \cite{Blaizot:2012fh,Blaizot:2013vha} recovering, as a particular case, the rate equations by the AMY group. The main limitation of these resummations and the corresponding implementation in the AMY approach, is that they apply to arbitrarily large medium length $L$, leading to an enhanced energy loss, while finite length effects are relevant in phenomenological implementations, in particular to avoid over-representation of the soft part of the spectrum. The main advantage of the quenching weights is that these finite length effects can be included, although, as mentioned, with no formal proof. So, we will use here the quenching weights \cite{Salgado:2003gb} as a theoretically motivated and phenomenological sound approach. In addition, this procedure is easy to implement,  allows to fit the quenching parameter $\hat q$ once the geometry of the medium is known and, moreover, it has been extensively tested, mainly for RHIC phenomenology \cite{Gyulassy:2001nm,Eskola:2004cr,Dainese:2004te,Armesto:2009zi} but also for the LHC, e.g.~\cite{Renk:2011gj}.

The formalism relies on two basic assumptions: i) the subsequent medium-induced gluon emissions are independent and ii) the fragmentation functions are not modified, i.e., fragmentation takes place in vacuum. These two assumptions find strong theoretical support in the recent analyses of coherence effects in the medium. Starting with the simplified setup of a QCD antenna \cite{MehtarTani:2010ma,MehtarTani:2011gf,CasalderreySolana:2011rz,MehtarTani:2012cy}, a pair of color-correlated partons with opening angle $\Theta$ emitting a  soft gluon, a simple picture of jet quenching arises \cite{CasalderreySolana:2012ef}: a medium of length $L$ and jet quenching parameter $\hat q$ has a typical transverse momentum scale for color correlations  $\Lambda_\perp\sim1/\sqrt{\hat qL}$; when the typical transverse size of the jet, $r_\perp\sim \Theta L$, is smaller than this scale, $r_\perp<\Lambda_\perp$, the medium cannot resolve the inner structure of the jet, which remains unchanged, but the whole jet radiates medium-induced gluons with the total charge of the jet. This is the {\it totally coherent case}. Clearly, while this picture indicates that the fragmentation function remains basically unmodified if color coherence is maintained, it still depends on the fraction of momentum $z$ and only a global energy loss affects the production of the fragmenting particles. This picture of jet quenching dictated by color coherence is in qualitative agreement with the experimental findings at the LHC \cite{Chatrchyan:2011sx,Chatrchyan:2012gw,Aad:2012vca,Chatrchyan:2013kwa,Chatrchyan:2014ava,Aad:2014wha} --- see \cite{Mehtar-Tani:2014yea} for a quantitative analysis of some data. 

Regarding the first assumption, as mentioned before, similar color coherence arguments can be used to prove that interference effects, which would break independent emission, are absent when the formation time of the medium-induced gluons, $\tau_{\rm form}\sim\sqrt{\omega/\hat q}$, is much smaller than the total length of the medium \cite{Blaizot:2012fh}. So, for soft radiation, $\omega\ll\omega_c\equiv \frac{1}{2}\hat q L^2$,  $\tau_{\rm form}\sim\sqrt{\omega/\hat q}\ll L$, and independent gluon emission is a good approximation. Notice that the quenching weights and the rate equations are equivalent for the case of soft radiation and when finite energy effects can be neglected, i.e., when the kernel depends neither on the energy of the parent parton nor on the medium length \cite{Turbide:2005fk}.

We use the quenching weights $P(\epsilon)$ tabulated in \cite{qw} to model the amount of energy loss of highly energetic partons (or better, coherent jets) which will eventually fragment in the vacuum to give a given hadron $h$. The corresponding cross section reads 
\begin{equation}
\frac{d\sigma^{AA\to h}}{dydp_T}=\int dq_T\,dz\frac{d\sigma^{AA\to k}}{dydq_T}\,P(\epsilon)\,D_{k\to h}(z,\mu_F^2)\,\delta\left(p_T-z(1-\epsilon)q_T\right),
\label{eq:crossec1}
\end{equation}
\noindent
where the cross section for producing parton $k$ is (we are neglecting here any difference from  parton to  hadron rapidities and we take all renormalization, factorization and fragmentation scales to be equal, $\mu_F=p_T$)
\begin{equation}
\frac{d\sigma^{AA\to k}}{dydq_T}=\int dx_1dx_2\,x_1f_i^A(x_1,\mu_F^2)\, x_2f_j^A(x_2,\mu_F^2)\frac{d\hat\sigma^{ij\to k}}{d\hat t}\, .
\label{eq:crossec2}
\end{equation}

\noindent
We make all calculations at NLO using the code in \cite{Stratmann:2001pb} with the proton PDF set CTEQ6.6M \cite{Nadolsky:2008zw}, the nuclear corrections to PDFs given by EPS09 \cite{,Eskola:2009uj} and vacuum fragmentation functions DSS \cite{deFlorian:2007aj,deFlorian:2007ekg}.

The quenching weights, $P(\epsilon)$, are defined as 
\begin{eqnarray}
  P(\epsilon) = \sum_{n=0}^\infty \frac{1}{n!}
  \left[ \prod_{i=1}^n \int dx_i \frac{dN(x_i)}{dx}
    \right]
    \delta\left(\epsilon - \sum_{i=1}^n x_i\right)
    \exp\left[ - \int_0^\infty \hspace{-0.2cm}
      dx \frac{dN}{dx}\right]\, ,
   \label{eq:spec1}
\end{eqnarray}
where the medium-induced radiation spectrum is given by\footnote{For simplicity, we omit here some technicalities related with a subtraction term, an interference between medium and vacuum radiation, and we refer the readers to the original papers.}
\begin{eqnarray}
  \omega\frac{dN}{d\omega d{\bf k_\perp}}
&=& {\alpha_s\,  C_R\over (2\pi)^2\, \omega^2}\,
    2{\rm Re} \int_{0}^{L}\hspace{-0.3cm} dy_l
  \int_{y_l}^{L} \hspace{-0.3cm} d\bar{y}_l\,
   \int d{\bf u}\,  
  e^{-i{\bf k}_\perp\cdot{\bf u}}   \,
  e^{ -\frac{1}{2} \int_{\bar{y}_l}^{L} d\xi\, n(\xi)\,
    \sigma({\bf u}) }\,\times\nonumber\\
&\times&  {\partial \over \partial {\bf y}}\cdot
  {\partial \over \partial {\bf u}}\,{\cal K}({\bf y}=0,y_l;{\bf u},\bar y_l) \, ,   \label{eq:spec2}
\end{eqnarray}
with the path integral
\begin{equation}
{\cal K}({\bf x},y;{\bf y},\bar y)\equiv\int_{{\bf r}={\bf x}(y)}^{{\bf r}={\bf y}(\bar y)} {\cal D}{\bf r} \exp\left\{i\frac{\omega}{2}\int d\xi \left[\frac{d{\bf r}}{d\xi}\right]^2-\frac{1}{2}\int d\xi\hat n(\xi)\sigma({\bf r})\right\}\, .
\label{eq:pathint}
\end{equation}

(Semi)analytical solutions of equations (\ref{eq:spec2}) and (\ref{eq:pathint}) are only known in a limited number of cases. Two main approximations are normally employed. One of them consists of expanding the exponents in series of the {\it opacity} parameter $n(\xi)\sigma(r)$ with a Fourier transformed cross section given by thermal QCD in some approximation, for example
\begin{equation}
\sigma({\bf q})=\frac{\mu^2}{q^2(q^2+\mu^2)}\ ,
\label{eq:hotcsect}
\end{equation}
with $\mu$ the thermal mass. The first term of this opacity expansion is normally employed, as used, for example in Ref. \cite{Gyulassy:2000er}.

The leading term of the Fourier transform of Eq. (\ref{eq:hotcsect}) is proportional to $r^2$ with a logarithmic correction. The second approximation consists of neglecting this logarithmic correction and approximate 
\begin{equation}
n(\xi)\sigma(r)\sim\frac{1}{2}\hat q(\xi)\, r^2.
\label{eq:qhatdef}
\end{equation}
With this approximation, the path integral is that of a harmonic oscillator with imaginary (eventually time-dependent) frequency. The solutions of this path integral when $\hat q(\xi)\sim 1/\xi^\alpha$ can be found in Ref. \cite{Salgado:2003gb}. This second approximation, sometimes known as the multiple soft scattering approximation, or BDMPS approximation, will be used in this paper. Eqs. (\ref{eq:pathint}) and (\ref{eq:qhatdef}) can be considered as our definition of the transport coefficient $\hat q$. 

The main difference between the two approximations, at the analytical level, is the presence of perturbative, power-law, tails in the opacity expansion, which are absent in the multiple soft scattering one. Notice that the AMY approach would correspond to a resummation of the multiple scatterings with the correct cross section (\ref{eq:hotcsect}) but without an interference between vacuum and medium radiation which turns out to be very relevant for finite medium lengths, making the soft part of the spectrum non-divergent when real angle emission is impossed for the emitted gluons.

\section{From a hydrodynamical profile to the transport coefficient}
\label{sect:hydroimp}

The distributions of the energy loss, $\Delta E$, of a fast quark or gluon, $i$, traversing a medium is computed through the quenching weights, $P_i(\Delta E/\omega_c,R)$ tabulated in \cite{qw,Salgado:2003gb} for the case of a static medium of finite length $L$ and transport coefficient $\hat q$, where
\begin{equation}
\omega_c=\frac{1}{2}\hat qL^2,\ \ R=\omega_cL\, .
\label{eq:omcR}
\end{equation}
For the expanding medium case, with proper time dependence of the transport coefficient, $\hat q(\tau)\sim 1/\tau^\alpha$, a dynamical scaling law was found \cite{Salgado:2002cd} that relates the resulting spectra with an equivalent static scenario. Based on this scaling law,  effective $\omega_c^{eff}$ and $R^{eff}$ for an hydrodynamical medium profile are computed as
\begin{eqnarray}
\omega_c^{eff}(x_0,y_0,\tau_{\rm prod},\phi)=\int d\xi\,\xi\,\hat q(\xi),\\
R^{eff}(x_0,y_0,\tau_{\rm prod},\phi)=\frac{3}{2}\int d\xi\,\xi^2\, \hat q(\xi)
\label{eq:omceff}
\end{eqnarray}
which, in particular, reproduce Eqs. (\ref{eq:omcR}) for the static case.
Similar implementations of the hydrodynamical model have been used before \cite{Armesto:2009zi,Burke:2013yra,Renk:2011gj,Bass:2008rv}, so it is a rather standard procedure\footnote{Notice that we have slightly changed the prescription to compute $R^{eff}$ which is now the second moment of $\hat q(\xi)$. The results are similar with the old prescription (see e.g. Eqs. (4.2)-(4.4) in Ref. \cite{Armesto:2009zi}) but with improved stability for functional dependences of $\hat q(\xi)$ that are divergent in $1/\xi$.}. The production point of the parton at time $\tau_{\rm prod}$ is distributed according to an $N_{\rm coll}$-scaling in the transverse plane and the azimuthal angle $\phi$ is taken as a random number in $[0,2\pi]$. As usual, each parton traverses the medium in a straight-line trajectory parametrized by the proper time $\xi$ at each point in the transverse plane. We only need to specify the relation between the local value of the hydrodynamical variables at $(x_\perp(\xi), y_\perp(\xi))$ and the local value of the transport coefficient $\hat q(\xi)$. Following our previous work \cite{Armesto:2009zi} we define
\begin{equation}
\hat q(\xi)=K\cdot 2\epsilon^{3/4}(\xi),
\label{eq:qhatlocal}
\end{equation}
where $K\simeq 1$ would correspond to the ideal QGP (see the estimate in e.g. Ref. \cite{Baier:2002tc}). Other relations between the transport coefficient and the local thermodynamical quantities have been explored e.g. in Ref. \cite{Bass:2008rv}. The local energy density $\epsilon(\xi)$ is taken from a hydrodynamical model of the medium, for which we will consider several different options in the next sections. The rest of the formalism follows that in Ref.  \cite{Armesto:2009zi}; we include here some details, while the complete formulation can be found in that reference.

In a dynamical medium like the one considered here, there is an ambiguity on the value of the transport coefficient, defined by Eq. (\ref{eq:qhatlocal}), for values smaller than the  proper time $\tau_0$ when relativistic hydrodynamics is started. One extreme case is to take $\hat q(\xi)=0$ for $\xi<\tau_0$. The absence of any energy-loss effect for these  early times is a strong assumption 
since neither thermalization nor isotropization is  necessary in the approach in which the quenching weights have been computed. 
To quantify this uncertainty, we consider three different extrapolations for the time from the hard parton production
to the thermalization:
\\
\noindent (i) $\hat q(\xi)=0$ for $\xi<\tau_0$;
\\
\noindent (ii) $\hat q(\xi)=\hat q(\tau_0)$ for $\xi<\tau_0$; and
\\
\noindent (iii) $\hat q(\xi)=\hat q(\tau_0)/\xi^{3/4}$ for $\xi<\tau_0$.
\\
These extrapolations range from the most extreme assumption of no effect at all before the thermalization time (case (i)) 
to a continuous interaction from the production time (taken to be $\tau_{\rm prod}\simeq 0.04$ fm/$c$) and a 
free-streaming medium with energy density dropping as $\epsilon(\xi)\sim1/\xi$ (case (iii)).

The production point of the hard scattering is characterized by a production weight $w(x_0,y_0)$ computed as
\begin{equation}
w(x_0,y_0)=T_A(x_0,y_0)T_A(\vec b-(x_0,y_0)),
\label{eq:prodpoint}
\end{equation}
where $T_A$ are the profile functions computed from a 3-parameter Fermi distribution at a given impact parameter $\vec b$ taken from \cite{DeJager:1987qc}. This weight allows to compute the average fragmentation functions for a parton $k$ which has traversed the medium and hadronizes in the vacuum to a hadron $h$ as
\begin{equation}
D^{\rm med}_{k\to h}(z,\mu_F^2)=\frac{1}{N}\int d\phi dx_0dy_0 w(x_0,y_0)\int \frac{d\zeta}{1-\zeta}P_k(x_0,y_0,\phi,\zeta)D^{\rm vac}_{k\to h}\left(\frac{z}{1-\zeta},\mu_F^2\right),
\label{eq:modFF}
\end{equation}
where $P_k(x_0,y_0,\phi,\zeta)$ is the quenching weight for parton $k$ and the normalization is $N=2\pi\int dx_0dy_0w(x_0,y_0)$. With this definition, the cross section, Eq.(\ref{eq:crossec1}), can be simply computed as

\begin{equation}
\frac{d\sigma^{AA\to h}}{dydp_T}=\int dq_T\,dz\frac{d\sigma^{AA\to k}}{dydq_T}\, D^{\rm med}_{k\to h}(z,\mu_F^2)\,\delta\left(p_T-zq_T\right).
\label{eq:crossec3}
\end{equation}

Let us emphasize again that the formalism assumes no medium modification of the fragmentation function, an early approximation which we have motivated here by color coherence arguments. For this reason, Eq. (\ref{eq:modFF}) should be understood as a suitable computational tool to include the energy loss of the coherent colored object traversing the QCD medium which eventually fragments in the vacuum. 

\section{Hydrodynamical model of the medium}
\label{sect:hydro}

We obtain the space-time distribution of the local energy density by solving the relativistic  hydrodynamic equations.   Such simulations require the specification of initial values for the energy momentum tensor, as well as parameters that describe medium properties, neither of which are accurately known.  In order to test the robustness of our results and conclusions with respect to these uncertainties, we repeat all calculations using space-time profiles from several different hydrodynamic simulations.

The first, which we refer to as ``Hirano'', corresponds to the calculation described in \cite{Armesto:2009zi,Hirano:2001eu,Hirano:2002ds,hydro-site}, to which we refer the reader for details.  In short, this calculation uses an optical Glauber model where the initial entropy density at initial proper time $\tau_0 = 0.6$ fm is given by a linear combination of the number density of participant nucleons, $\rho_{\rm part}$, and binary collisions, $\rho_{\rm bin}$:
\begin{equation}
s \propto (1-x)\rho_{\rm part} + x \rho_{\rm coll},
\end{equation}
with binary collision fraction $x = 0.15$.  A bag model equation of state is used, with chemical freeze out enforced at $T_{ch} = 170$ MeV, and kinetic freeze out at $T_f = 100$ MeV, below which temperature the medium has frozen out and no energy loss occurs.  This is an ideal hydrodynamic calculation, with vanishing viscosity.

The other two hydrodynamical models correspond exactly to the calculations in \cite{Luzum:2008cw} (for 200 GeV Au-Au collisions at RHIC) and \cite{Luzum:2009sb} (for 2.76 TeV Pb-Pb collisions at the LHC), to which we again refer the reader for all relevant details. 

One calculation, which we refer to as ``Glauber'', uses for an initial condition an energy density proportional to the density of binary collisions, $\rho_{\rm bin}$, while the ratio of shear viscosity to entropy density is fixed to a constant value of $\eta/s = 0.08$.

The final calculation is referred to as ``fKLN''.   This simulation takes its initial condition from a factorised Kharzeev-Levin-Nardi model \cite{Drescher:2006pi}, with the shear viscosity set to $\eta/s = 0.16$.

Both of the latter simulations begin at an initial proper time of $\tau_0 = 1$ fm and use an equation of state inspired by lattice QCD calculations.  Each system is assumed to be in chemical equilibrium until it reaches a freeze out temperature of $T_f = 140$ MeV.

All of these calculations have been successfully tested against various experimental data, but use different choices for initial conditions, thermalization time, viscosity, equation of state, etc.  Thus, we expect that the variation in our results from using these different models should give a reasonable indication of the uncertainty coming from the hydrodynamic background.  We will see that such uncertainty is negligible with respect to our main conclusions.
\section{Results of the fits}
\label{sect:results}

We restrict our study here to the case of one-particle inclusive suppression at RHIC \cite{Adare:2008qa} and the LHC \cite{Abelev:2012hxa}, i.e., we stick to the simplest observable $R_{AA}$. We do not consider here other observables previously considered in fits like this, see e.g.~\cite{Armesto:2009zi}, as they may involve other effects related with fragmentation, mass effects on the energy loss mechanism, etc.
Something new is, however, the centrality dependence of both RHIC and LHC data. Most of previous analyses have only studied the most central class or analyzed the centrality dependence only for one energy \cite{Burke:2013yra,Renk:2011gj,Bass:2008rv}. 

\begin{figure}
\includegraphics[width=\textwidth]{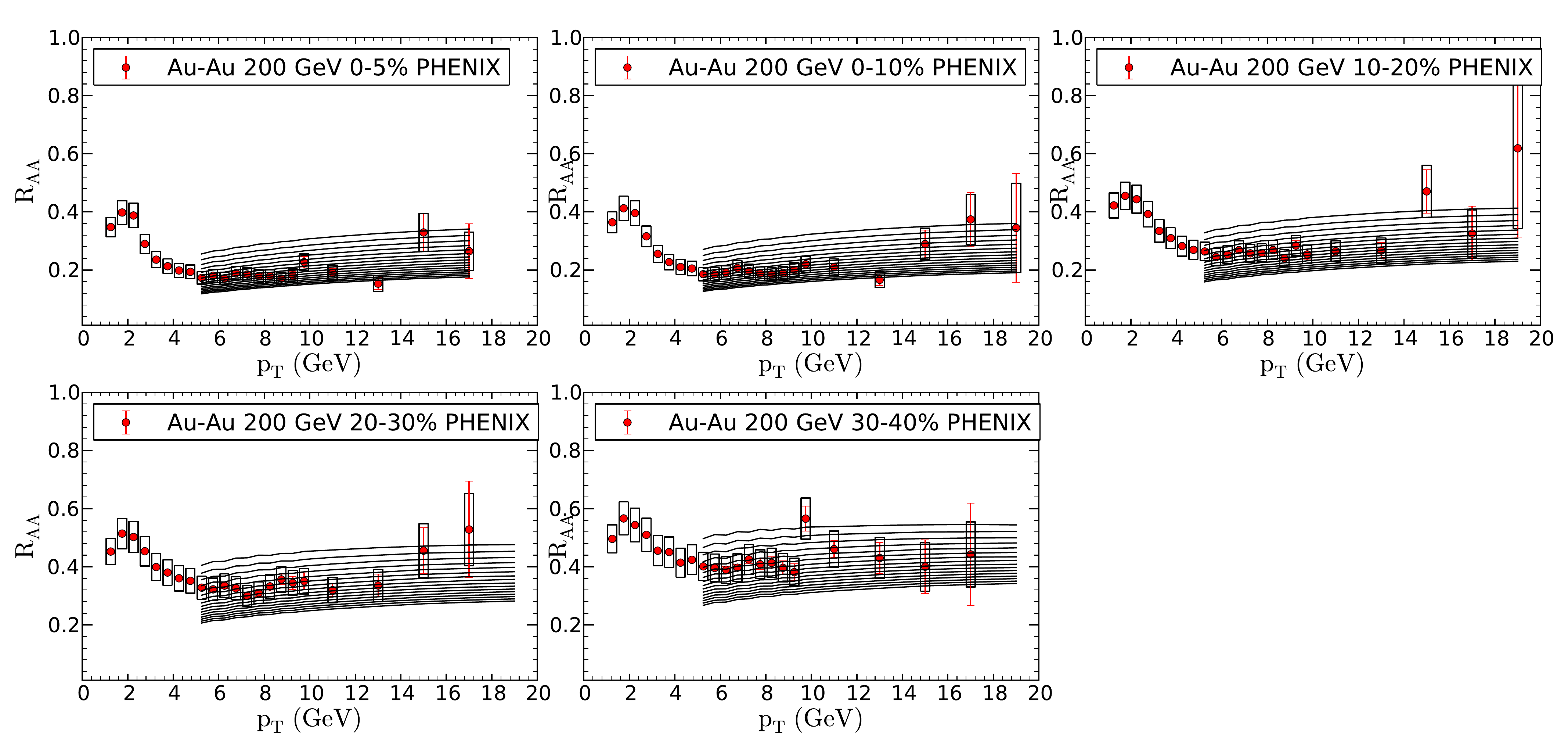}
\caption{Suppression of inclusive $\pi^0$ in AuAu collisions at $\sqrt{s_{\rm NN}}=200$ GeV for different values of the parameter $K$ (see Eq. (\protect\ref{eq:qhatlocal})) compared with PHENIX data at different centralities \cite{Adare:2008qa}. Curves from top to bottom correspond to $K=K^\prime/1.46$, with $K^\prime=2,2.25,2.5,\dots,6$, using the ``Hirano'' hydrodynamical model and the energy density prior to the start of hydrodynamical evolution taken as constant, see the previous Sections.}
\label{fig:raaRHIC}
\end{figure}

\begin{figure}
\includegraphics[width=\textwidth]{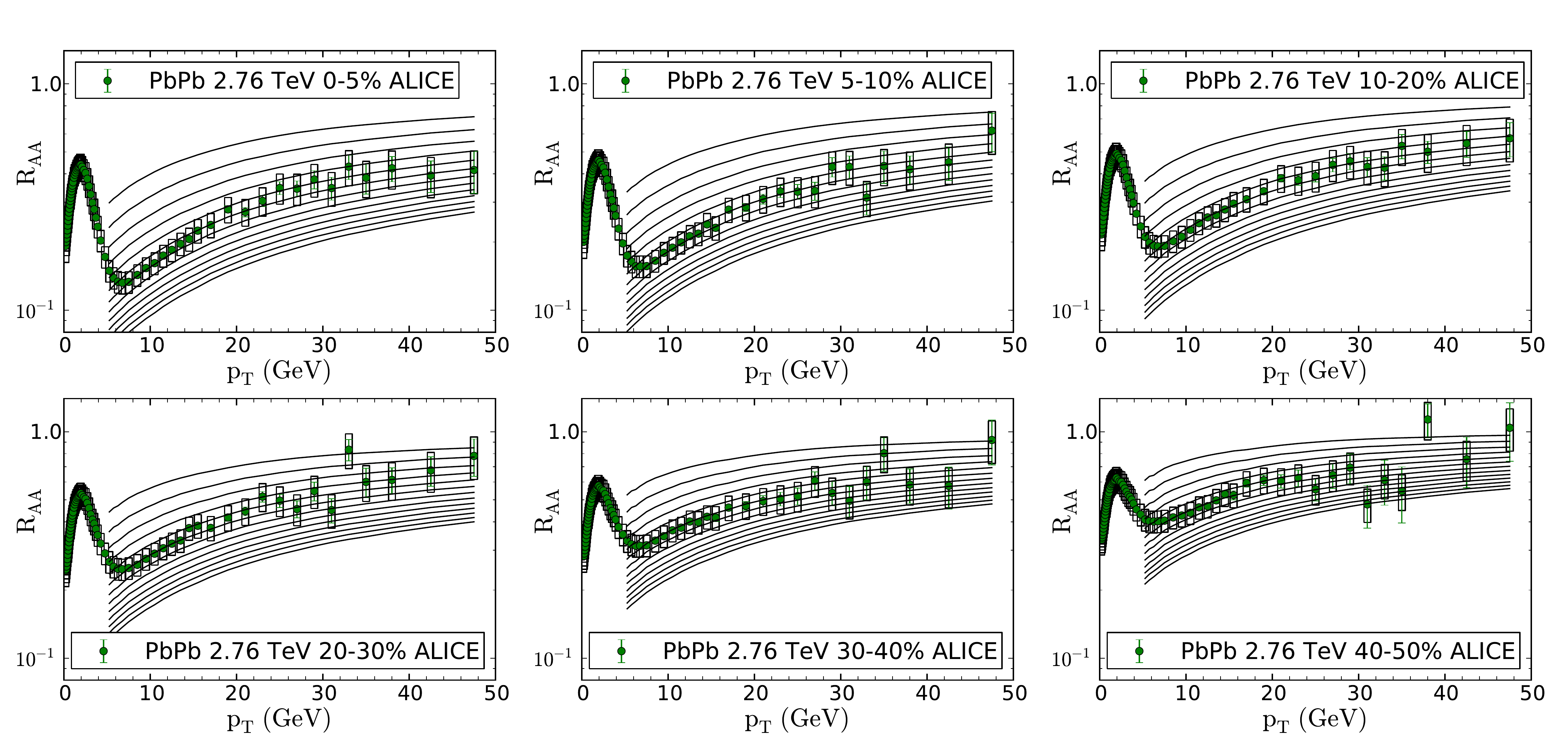}
\caption{Suppression of inclusive charged particles in PbPb collisions at $\sqrt{s_{\rm NN}}=2.76$ TeV for different values of the parameter $K$ (see Eq. (\protect\ref{eq:qhatlocal})) compared to ALICE data at different centralities \cite{Abelev:2012hxa}. Curves from top to bottom correspond to $K=K^\prime/1.46$, with $K^\prime=0.5,0.7,0.9,\dots,3.1$, using the ``Hirano'' hydrodynamical model and the energy density prior to the start of hydrodynamical evolution taken as constant, see the previous Sections.}
\label{fig:raaLHC}
\end{figure}

We have performed our study using three different hydrodynamical profiles and three different assumptions for the time prior to the equilibration time, see the previous Sections. For economy of space we present here only a subset of the total results which, anyway, lead to the same qualitative conclusions (only the actual value of the parameter $K$ depends on the given prescription). In Fig.~\ref{fig:raaRHIC} we plot our results for different values of $K$ together with the experimental data from the PHENIX Collaboration \cite{Adare:2008qa} on suppression of inclusive neutral pions on AuAu collisions at $\sqrt{s_{\rm NN}}=200$ GeV (72 data points). In Fig.~\ref{fig:raaLHC} we plot the corresponding results for LHC PbPb collisions at $\sqrt{s_{\rm NN}}=2.76$ TeV from ALICE \cite{Abelev:2012hxa}, 156 data points (notice that the plotted values of $K$ are different in both figures). We restrict to $p_T>5$ GeV/c to stay in a region where pQCD can be applied and no large contribution from other effects like flow is expected.

\begin{figure}
\includegraphics[width=0.9\textwidth]{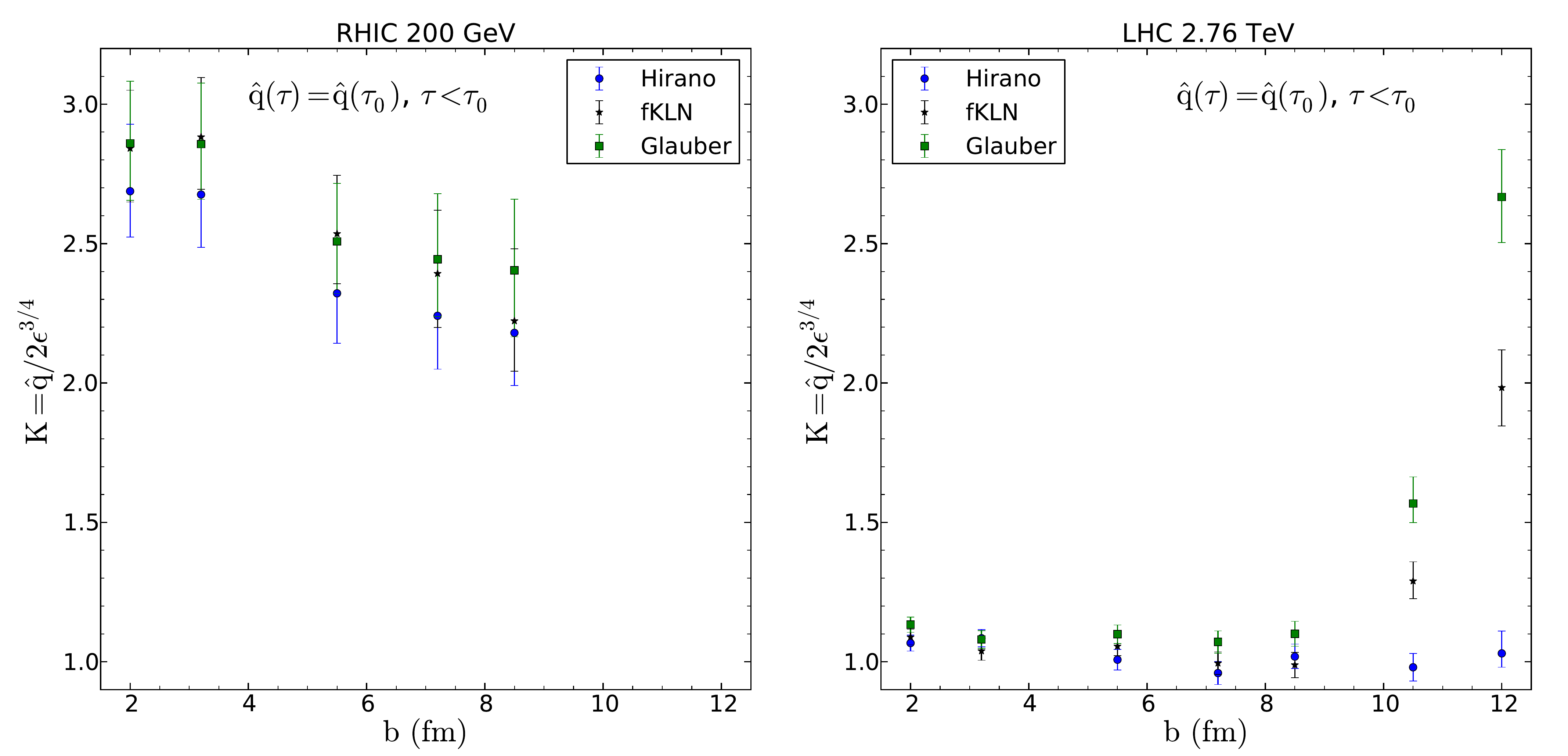}
\caption{$K$-factors obtained from fits to PHENIX $R_{\rm AA}$ data \cite{Adare:2008qa} \textit{(left panel)} and to ALICE $R_{\rm AA}$ data \cite{Abelev:2012hxa} \textit{(right panel)} using different hydrodynamical profiles as a function of the average impact parameter for each centrality class and the energy density prior to the start of hydrodynamical evolution taken as constant, see the previous Sections.}
\label{fig:chi2RHICandLHC}
\end{figure}

\begin{figure}
\includegraphics[width=0.9\textwidth]{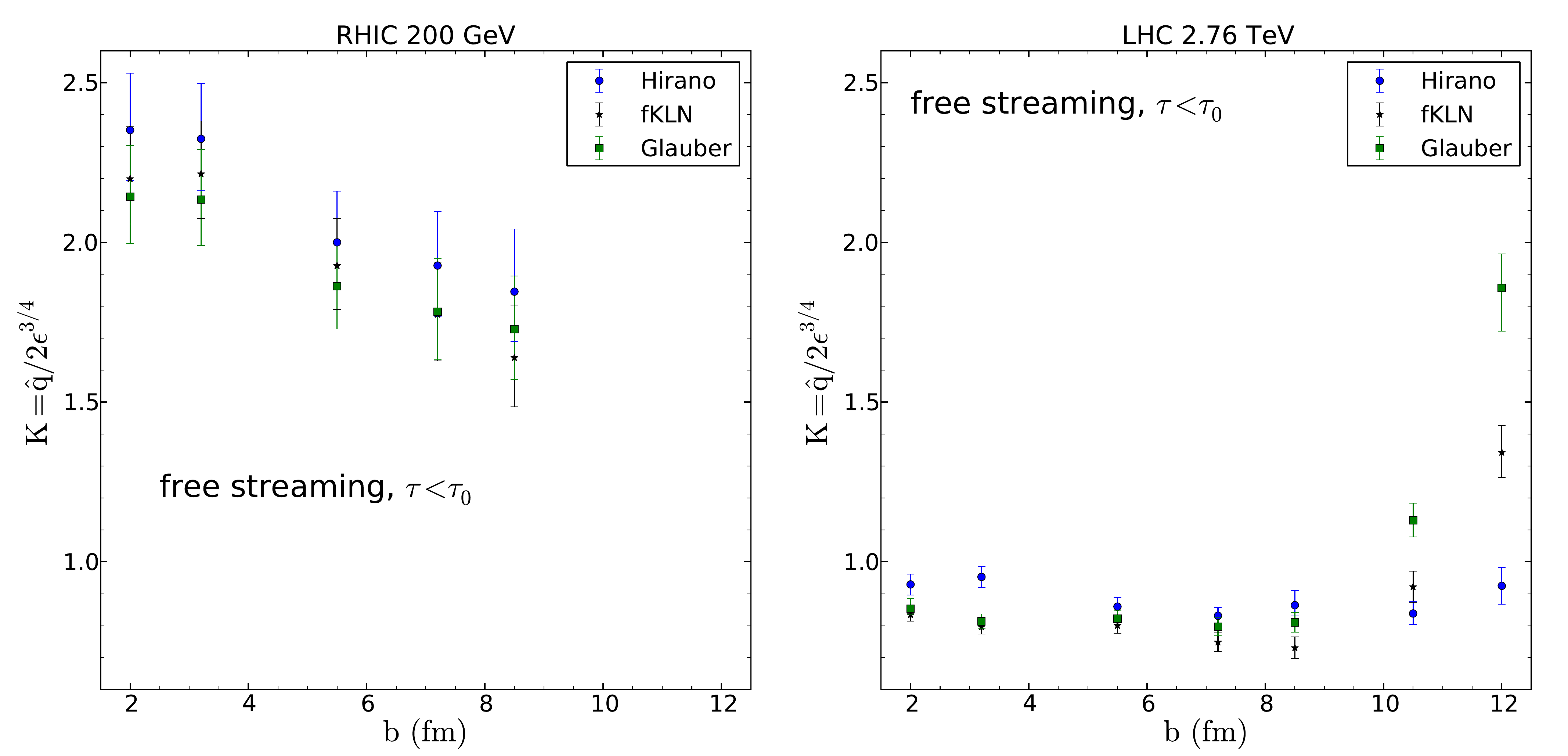}
\caption{$K$-factors obtained from fits to PHENIX $R_{\rm AA}$ data \cite{Adare:2008qa} \textit{(left panel)} and to ALICE $R_{\rm AA}$ data \cite{Abelev:2012hxa} \textit{(right panel)} using different hydrodynamical profiles as a function of the average impact parameter for each centrality class and for the free-streaming extrapolation, see the previous Sections.}
\label{fig:chi2RHICandLHCfreestreaming}
\end{figure}

\begin{figure}
\includegraphics[width=0.9\textwidth]{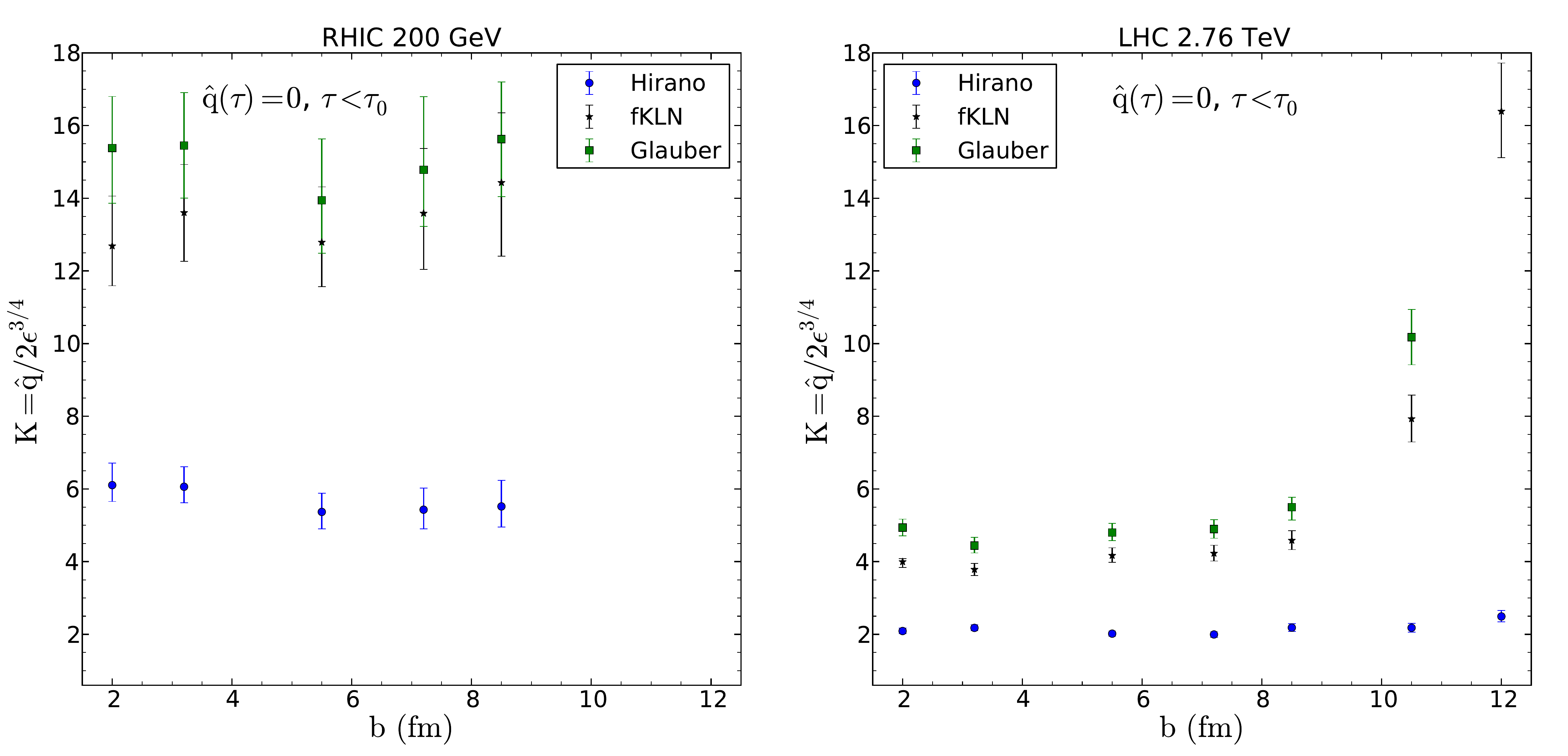}
\caption{$K$-factors obtained from fits to PHENIX $R_{\rm AA}$ data \cite{Adare:2008qa} \textit{(left panel)} and to ALICE $R_{\rm AA}$ data \cite{Abelev:2012hxa} \textit{(right panel)} using different hydrodynamical profiles as a function of the average impact parameter for each centrality class and for $\hat q(\xi)=0$ before thermalization, see the previous Sections.}
\label{fig:chi2RHICandLHCqhat0}
\end{figure}

We have performed a $\chi^2$ fit to the best value of $K$ for each energy and centrality, and for each assumption of hydrodynamical profile or behavior of $\hat q$ at values of proper time smaller than the thermalization time $\tau_0$ assumed in each hydrodynamical simulation. For the case of ALICE data \cite{Abelev:2012hxa} we add the systematic and statistical errors in quadrature, as no particular instructions of how to include them in a fit are provided. For the case of RHIC, the latest analysis includes the contribution from several different error sources. The two methods lead to comparable values of $K$ (differences $\sim 5$\%) except for the most peripheral bins, for which the $K$ values in the case of errors added in quadrature are $\sim 30$\% smaller. The uncertainty band is determined by $\Delta \chi^2 =$ 1. In order to make the comparison between RHIC and the LHC, these issues need to be taken into account, although the conclusions do not change at the qualitative level. In the left panels of Fig.~\ref{fig:chi2RHICandLHC}, Fig.~\ref{fig:chi2RHICandLHCfreestreaming} and  Fig.~\ref{fig:chi2RHICandLHCqhat0} we plot the different values of the $K$-parameter fitted to the PHENIX data \cite{Adare:2008qa} for different combinations of hydrodynamical profiles and behavior before the thermalization time. The corresponding values for the LHC \cite{Abelev:2012hxa} are plotted in the right panels of the same figures.

Several comments are in order. First, the extracted values of $K$ are compatible for the cases of either frozen energy density or free streaming before $\tau_0$, and the results for the three different hydrodynamical implementations are similar. This is not the case when no quenching is assumed before $\tau_0$; for this assumption, the two viscous hydrodynamic models that use a common (larger) $\tau_0$ require a larger $K$ than the ideal hydrodynamic model that considers a smaller $\tau_0$, with actual values which become unrealistically large. Therefore, we do not consider the results obtained for this assumption for the discussion of the values of $K$, but the qualitative behavior that we find is in agreement with the two other assumptions.  In any case they  clearly illustrate the importance of the treatment of early times in jet quenching computations. Second, for the most peripheral collisions at the LHC,  model ``Glauber"  demands a much larger $K$ than the others, while model ``Hirano"  returns a rather flat value of $K$ for all centralities. Third, the trend of the results at RHIC is a slight decrease with decreasing centrality, although compatible with constant, while at the LHC the behavior is constant except for the smaller centralities, where the behavior, as it was mentioned above, depends very much on the hydrodynamical profile employed.

\begin{figure}
\includegraphics[width=\textwidth]{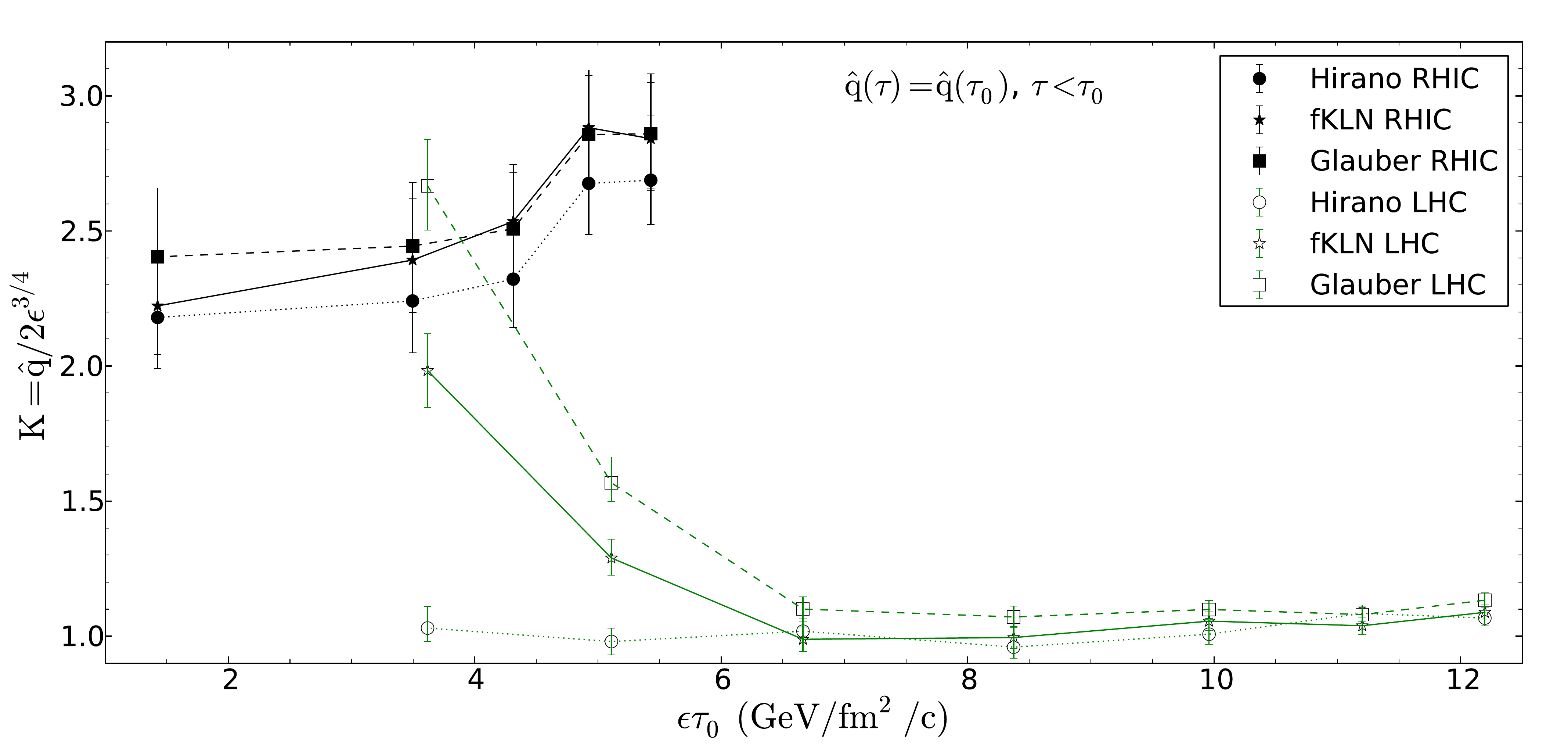}
\caption{$K$-factor obtained from fits to $R_{AA}$ data at RHIC and LHC energies for different centrality classes plotted as a function of an estimate of the energy density times formation time $\tau_0$ of the QCD medium formed in each case. The $\epsilon\tau_o$ estimates are taken from Refs. \cite{Adare:2015bua,Adam:2016thv}.}
\label{fig:overlapRHIC-LHC}
\end{figure}

In the end, we would like to understand the systematics and relation of LHC and RHIC results for the $K$-factor that we obtain. First, we notice that, in principle, Eq. (\ref{eq:qhatlocal}) determines how far or close the perturbative estimate $\hat q\simeq 2\epsilon^{3/4}$ is from our value fitted to experimental data. In this sense, we note that there is a clear departure from unity of this value for the case of RHIC. This fact was found several times \cite{Armesto:2009zi,Bass:2008rv}\footnote{Note that the difference of the present extraction $K\sim$ 2--3 and $K\sim 4$ from \cite{Armesto:2009zi} comes mainly from the new definition of $R$ in (\ref{eq:omceff}), indicating, again, the important role of the geometry in the extraction of $\hat q$.}. We also find that the corresponding value of $K$ is smaller at the LHC, a fact which has been already found before by other groups \cite{Burke:2013yra} but with a smaller decrease (a factor $\sim 25$ \% compared to our factor 2--3)\footnote{Nevertheless this comparison needs to be taken with caution as the values of $\hat q/T^3$ quoted in \cite{Burke:2013yra} are performed at a given temperature and no systematics with temperature is presented. Moreover,  $\hat q$ is not the natural fitting parameter in the models studied in that reference but a derived quantity once the parameters of the different models are extracted from the data. In our case, $\hat q$ is the natural parameter, given by Eqs. (\ref{eq:pathint}) and (\ref{eq:qhatdef}), and the $K$ factor has a well defined meaning.}. The study of the centrality dependence is, nonetheless, more surprising. The extracted value of $K$ seems to depend mainly on the energy of the collision and much less (if any) on the centrality. This is not the behavior one would expect from a naive interpretation in which the $K$ factor only indicates the departure from the leading order perturbative estimate determined by temperature. In this naive interpretation, a medium with a smaller temperature (RHIC) would need higher orders of the perturbative series to be included, while a medium at higher temperature would be closer to the ideal limit. This simple interpretation does not correspond, notwithstanding, to the present findings as there is an overlap on typical energy densities between central AuAu at RHIC and semi-peripheral PbPb at the LHC, so their values of $K$ should coincide in this naive interpretation. In order to provide an estimate of this overlap, we plot, in Figure \ref{fig:overlapRHIC-LHC}, the $K$-factors obtained for different centralities and energies versus an energy density times formation time $\tau_0$ extracted from the experimental data using Bjorken estimates \cite{Adare:2015bua,Adam:2016thv} --- we have checked that the overlap is similar if we plot as a function of the maximum energy density of the hydrodynamical profiles that we have used to perform the fits. 

\begin{figure}
\includegraphics[width=\textwidth]{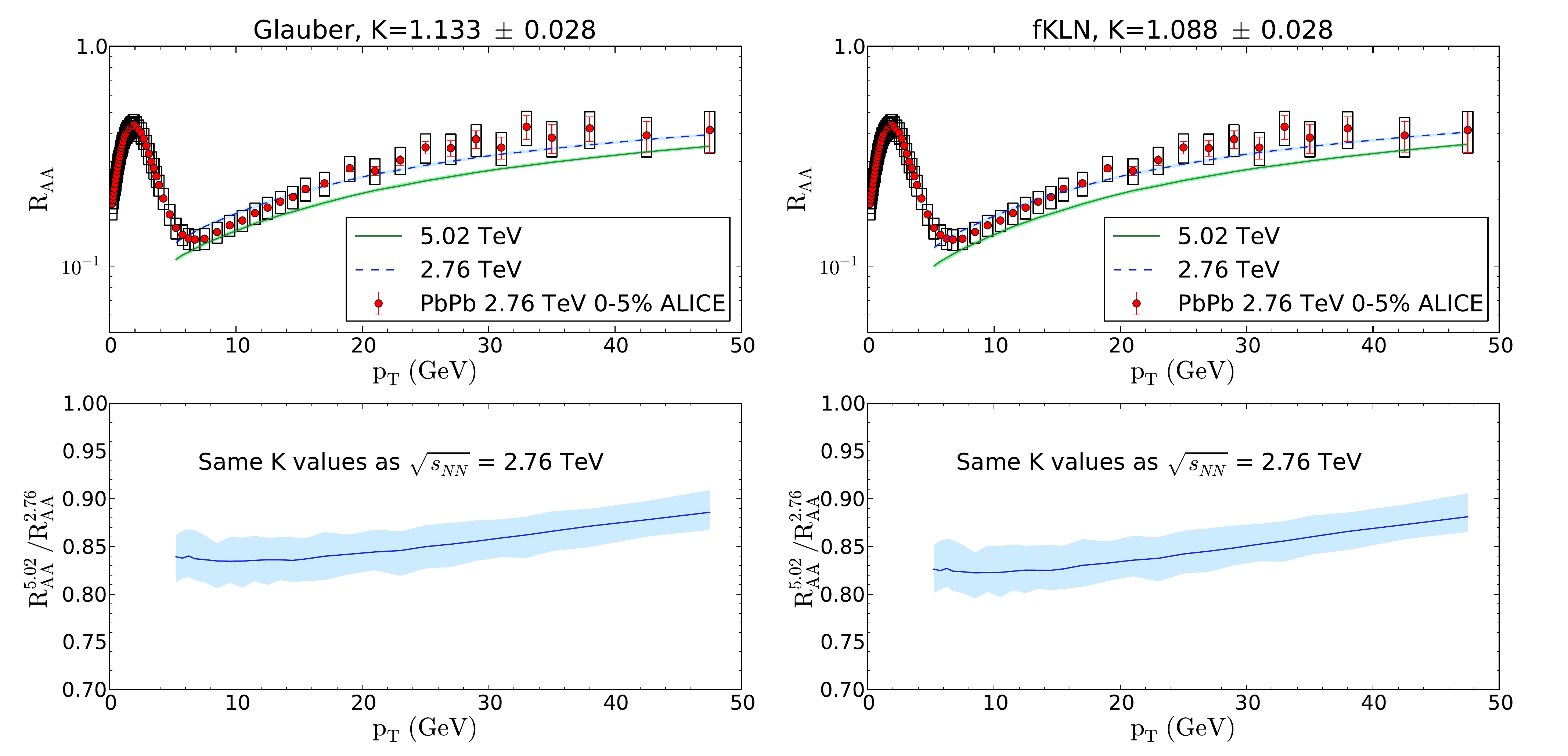}
\caption{Top: Curves for PbPb collisions at $\sqrt{s_{\rm NN}}=2.76$ (dashed blue) and 5.02 (solid green) TeV and the 0-5\% centrality class using ``Glauber'' \textit{(left panel)} and ``fKLN'' \textit{(right panel)} hydrodynamical evolution and the energy density prior to the start of hydrodynamical evolution taken as constant, see the previous Sections. ALICE $R_{AA}$ data at $\sqrt{s_{\rm NN}}=2.76$ TeV for the same centrality class \cite{Abelev:2012hxa} are also shown. Bottom: Ratios of the corresponding curves for 5.02 TeV w.r.t. 2.76 TeV. The used values of $K$ are shown above the top plots.}
\label{fig:LHCprediction}
\end{figure}

Finally, we would like to include here the predictions of our formalism for the forthcoming data on PbPb collisions at the LHC at $\sqrt{s_{\rm NN}}=$5.02 TeV. In our case, the $K$-factor is not fixed but fitted from experimental data.  However, assuming that the perturbative estimate $\hat q\simeq 2\epsilon^{3/4}$ is approximately correct, it would be reasonable to think that our $K$-factor cannot be much smaller than unity. So, assuming the same values of $K$ as the ones we obtain from the fit to $\sqrt{s_{\rm NN}}=$2.76 TeV data, $K$=1.133 $\pm$ 0.028 for ``Glauber'' hydrodynamic profile and  $K$=1.088 $\pm$ 0.028 for ``fKLN'' hydrodynamic profile and considering a frozen $\hat q$ between production time and $\tau_0$, we obtain a slightly stronger suppression for the 5.02 TeV case, see Fig.~\ref{fig:LHCprediction}. (Results from the ``Hirano'' ideal hydrodynamic model are not available for $\sqrt{s_{\rm NN}}=$5.02 TeV, but we note that the prediction for the \textit{change} in $R_{\rm AA}$ with collision energy is essentially independent of the hydro model). 

This behavior is easy to understand considering that the transverse momentum slope is not that different at the two LHC energies, but the larger energy densities at the higher collision energy imply larger absolute values of $\hat q$ for the same $K$-factor. Thus, the energy loss is larger and the suppression becomes stronger. The effect is, however, not very large. 

\section{Conclusions}

We have studied the one-particle inclusive suppression of particles produced at high transverse momenta at RHIC and the LHC as a function of centrality. By defining a constant $K$-factor  with respect to the perturbative estimate $\hat q\simeq 2\epsilon^{3/4}$ we fit the corresponding experimental data at RHIC and LHC for different centralities. The fitted value at RHIC confirms previous estimates \cite{Armesto:2009zi,Bass:2008rv} of large corrections to the ideal case, although the actual numerical value is a bit smaller, due to a new, more stable, definition of the effective values of the static scenario equivalent to the evolving medium, Eq.  (\ref{eq:omceff}). For the case of the LHC, instead, the extracted value of $K$ is close to unity. One would be tempted to make the naive interpretation that the medium created at the LHC, having a larger temperature, is closer to the ideal case than the one at RHIC, for which larger corrections or even a strongly coupling treatment, could be needed. 

This naive interpretation finds difficulties to be accommodated, however, with the fact that the centrality dependences at RHIC and the LHC separately are rather flat, that is, the change in the value of $K$ is not simply due to the different temperature (or energy density), as there is a large region of overlap between RHIC and the LHC for different centralities. At this moment we do not have an interpretation for this finding which, in any case, should be checked by other model implementations of jet quenching. It is also worth noticing that the extraction of the value of $K$ in the case of RHIC depends on a single set of experimental data, namely inclusive $\pi^0$ suppression measured by PHENIX. The corresponding results from STAR on $\pi^++\pi^-$ suppression \cite{Agakishiev:2011dc} show a smaller suppression but the smaller range of transverse momentum studied makes our analysis to be not very reliable. For this reason we have chosen not to include this set of data in the fit. For the LHC, on the other hand, CMS \cite{CMS:2012aa} and ATLAS \cite{Aad:2015wga} have measured the suppression of inclusive charged particles with results almost identical to the ones from the ALICE collaboration\footnote{On the other hand, ALICE data are restricted to midrapidities where the boost invariant picture of the medium underlying the initial conditions for the hydrodynamic calculations should hold with very good accuracy, while ATLAS and CMS cover a much wider rapidity region. Further difficulties come from the modelling of the energy loss far from midrapidity. For these reasons, we restrict our study at the LHC to ALICE data.}. 

From a theoretical perspective, the formalism of the quenching weights presents several limitations which could influence the result. We quote some of them here: (i) the very definition of $\hat q$ in the path integral, Eq. (\ref{eq:pathint}), neglects the perturbative tails of the distributions which may enhance the energy loss and even change its angular dependence; (ii) as we have mentioned, the quenching weights rely on two assumptions which could fail if color coherence of the parton shower is broken during the path of the jet though the medium; (iii) the geometrical implementation of the hydrodynamical profiles relies on the relations (\ref{eq:omceff}) which have been proven only for a class of profiles $\hat q(\tau) \propto 1/\tau^\alpha$; (iv) finite length corrections to the independent gluon emission are not known in any of the implementations used at present; (v) finite energy corrections to the medium-induced gluon radiation could also affect the result; (vi) the jet quenching parameter $\hat q$ is taken to be energy or length independent, while evolution equations have been proposed \cite{CasalderreySolana:2007sw,Blaizot:2014bha,Iancu:2014kga}; (vi) finite energy corrections could also contain {\it collisional energy loss} which is neglected in our formalism and which may have a different parametric dependence with the medium properties. In spite of these limitations, it is difficult to imagine how a more refined implementation of the in-medium parton shower could qualitatively modify our finding of a mostly flat, in centrality, value of $K$ and different for different collision energies.  At the level of the partonic spectra, the main quantity affecting the suppression is a decreasing value of the slope with increasing energy: for a simple parametrization of $1/p_T^\delta$, $\delta$ varies in the range $5-7$ from the LHC to RHIC. This steeply falling spectrum introduces a bias in the proved energy loss distributions, so the typical energy of the partons at the LHC is larger than at RHIC for the same measured $p_t$. In this way, a softer part of the energy loss distribution is probed with increasing $\delta$, so that the perturbative tails neglected in the multiple soft scattering approximation used may become relevant. Moreover, the typical jet could also be more collimated at lower energies (as the typical fraction of momentum probed in the fragmentation function is also larger).  These details could modify the value of $K$ extracted for a more realistic treatment of  jet coherence and thermal cross sections. A better control on the initial times and the study of different experimental observables with refined methods will allow to clarify this issue. 

As an outlook, we plan to study the effect of event-by-event fluctuations as presently done e.g. in \cite{Renk:2011qi,Zhang:2012ik,Noronha-Hostler:2016eow}.

\noindent {\bf Note added:} When finalizing the preparation of this manuscript, new, preliminary, experimental data on inclusive charged particle suppression $R_{AA}$ of PbPb collisions at $\sqrt{s}=5.02$ TeV measured by the CMS collaboration were presented at the 3rd International Conference on the Initial Stages in High-Energy Nuclear Collisions {\it (InitialStages2016)} in Lisbon (Portugal) \cite{InitialStages2016}. Although we have not included CMS data in our analysis, for the reasons explained above, taking at face value, the new data would indicate a suppression which is almost energy independent at the LHC. In Fig.~\ref{fig:LHCprediction} we predict a $\sim 15\%$ larger suppression at $\sqrt{s}=5.02$ TeV compared to $\sqrt{s}=2.76$ TeV. We have checked that, if confirmed, this same suppression would indicate that the $K$ value needed to reproduce the higher energy LHC data is $\sim 10\%$ smaller than the ones quoted here. Obviously, the exact value would require a new fit once the data for central rapidities are available.

\section*{acknowledgements}
We thank Yan Zhu for her participation in the early stages of this work and Guilherme Milhano, Risto Paatelainen, Jorge Casalderrey-Solana and Konrad Tywoniuk for discussions. We also thank Constantin Loizides, Abhijit Majumder, Jacquelyn Noronha-Hostler, Guang-You Qin, Krishna Rajagopal and Xin-Nian Wang for comments on the first version of the manuscript, and Tetsufumi Hirano for providing hydrodynamical profiles for the LHC. This research was supported by the European Research Council grant HotLHC ERC-2011-StG-279579; the People Programme (Marie Curie Actions) of the European Union's Seventh Framework Programme FP7/2007-2013/ under REA grant agreement \#318921 (NA); Ministerio de Ciencia e Innovaci\'on of Spain under project FPA2014-58293-C2-1-P; Xunta de Galicia (Conseller\'{\i}a de Educaci\'on) --- the group is part of the Strategic Unit AGRUP2015/11. C. Andr\'es thanks the Spanish Ministery of Education, Culture and Sports for financial support (grant FPU2013-03558).

\end{document}